\documentclass[12pt]{article}

\usepackage{fullpage}
\usepackage{epsfig}
\usepackage{amsfonts}
\usepackage{amssymb}
\usepackage{amsmath}

\newcommand{\mean}[1]{\langle #1 \rangle}

\newcommand{\fourierv}{\sum_{n=-\infty}^{\infty} v_n(t)e^{i n x}}
\newcommand{\fourierrho}{\sum_{n \neq 0} \rho_n(t)e^{i n x}}

\title{Flocking Regimes in a Simple Lattice Model}

\author{J. R. Raymond and M. R. Evans\\ 
{\small SUPA, School of Physics, University of Edinburgh,\newline
Mayfield Road, Edinburgh EH9 3JZ, UK}}

\begin{document}

\maketitle

\begin{abstract}

We study a one-dimensional lattice flocking model 
incorporating all three of the flocking criteria proposed by
Reynolds [{\em Computer Graphics} {\bf 21} 4 (1987)]: alignment, centring and separation.
The model generalises that introduced  by O. J. O' Loan and 
M. R. Evans
{\it J.  Phys.  A.} {\bf 32} L99 (1999).
We motivate the 
dynamical rules  by microscopic sampling considerations.
The model exhibits various flocking regimes: the alternating flock,
the homogeneous flock and dipole structures. We investigate these regimes numerically and within a continuum mean-field
theory.

\end{abstract}


\noindent {\bf Pacs:} 05.70.Fh  64.60.Cn  87.10e

\section{Introduction}
\label{introduction}

Flocking  refers to a family of behaviours regularly
seen in nature, including schooling of fish\cite{BDF} and flocking
of birds and  moths\cite{LG}.  Whilst the features that are truly essential to
flocks remain open to speculation\cite{Rey,JLM,OSM,LPP} it is clear
that the variety of biological\cite{HBSM,BDF,LG} and
engineered\cite{Cro} systems in which flocking behaviour does, or
could, prove useful is vast.  Statistical mechanics has recently
played an important role in uncovering and determining the source of
many flocking phenomena\cite{TTR}.

One of the pioneers in flocking research was Reynolds\cite{Rey}.
Within his work he determined three behaviours characteristic of
flocking: the desire to match the velocity of flockmates (alignment),
the desire to stay close to flockmates (centring), and the desire to
avoid collisions (separation).  Although Reynolds' original research
was motivated by applications in computer graphics, his definitions
have been adopted widely and implemented by  researchers
from  a variety of disciplines. Reynolds coined the term {\it boids} to
describe the generic self-propelled particles obeying these rules,
and we shall continue that convention in this paper.

  Within the physical science literature research has focused
primarily on the steady state behaviour of flocking models and the
non-equilibrium elements which cause solutions to differ from
equilibrium systems.  Many models involving only an alignment
interaction have been shown to demonstrate a symmetry broken velocity
state\cite{VCBCS}.  Results have also been found which are strongly
non-equilibrium in one and two dimensions, where equilibrium systems
do not undergo comparable symmetry breaking phase transitions\cite{TT2}.
Important work on alignment has included that of Vicsek et
al\cite{VCBCS} and Toner and Tu\cite{TT} who shed light on the
non-equilibrium processes involved.  A wide variety
of models also exist in two or three dimensions incorporating one or
several of Reynolds' rules besides alignment\cite{SSMHS,BDG}. The
lattice model of \cite{BDG}, which includes separation and alignment,
shows a density-dependent symmetry broken state, for example.  Similar
results have also been derived using a network approach to flocking,
whereby boids are nodes within a dynamic graph\cite{AH}.

There are fewer obvious physical realisations of one-dimensional
flocking, although analogy may be made with ring shaped aquariums, and
with pedestrian dynamics in corridors \cite{VCFH}.  However, research into
one-dimensional systems certainly is interesting from a fundamental
viewpoint: in particular, nonequilibrium phase transitions may be
exhibited\cite{Evans}.  In the one dimension alignment model of Czirok {\rm et
al}\cite{CBV} and the one dimensional lattice model of O'Loan and
Evans\cite{EOL} novel states not anticipated by higher dimensional
treatments were found.  In \cite{CBV} a symmetry broken steady state
was reported, whilst \cite{EOL} found a steady state which alternated its
direction of motion on an unusually short time scale $O(\ln N)$ where
$N$ is the number of boids in the system.  In the one dimensional model of Levine et
al\cite{LR} the effects of centring and separation were considered and
new behaviours were also found.

Our aim in this work is to introduce and study a one-dimensional
lattice model involving all three of Reynolds' flocking criteria.  We
do this by generalising the model of \cite{EOL}.  We demonstrate a
variety of new flocking regimes: further to the alternating flock
state found in \cite{EOL} we show that on small systems a homogeneous
flock may occur; also, the inclusion of centring may produce various
dipole structures in which the velocities of the boids point towards the centre
of the structure.  Our generalisation also addresses some
deficiencies in the dynamics of \cite{EOL} which implied full knowledge
of neighbouring boids---in the present work the dynamics is motivated
from the sampling of a finite number of neighbouring boids and yields
a smoother form [Eq. (\ref{abias}) below] for the expected update velocity of a
boid as a function of the average neighbourhood velocity.

The paper is organised as follows. In section 2 we define and motivate
the model to be studied. In section 3 we present numerical simulations
of the model and examine the different regimes that may be
observed. In section 4 we employ mean-field theory to provide further
evidence for these regimes.  We conclude with a discussion in section
5.

\section{Boid Dynamics}
\label{Model}
\subsection{Model synopsis}
\label{BDMS}

We first review the model of \cite{EOL} which consists of $N$ boids
inhabiting a lattice of $L$ sites with unit lattice spacing. 
The global density, $\mean{\rho}$,
is defined as $\mean{\rho} = N/L$.  Each boid $\mu$ is defined by a position
$x_\mu = 1\ldots L$ and velocity $v_\mu = \pm 1$.  An update of the system
consists of the following steps:

\begin{enumerate}
\item{A boid is chosen at random.}
\item{A {\it preferred direction} $U(x_\mu,t)= \pm 1$ or $0$ is determined from the velocity and spatial distribution of neighbouring boids.}
\item{$v_\mu$ is updated to $\pm 1$ with probability $W_{\pm 1}$}
  \begin{eqnarray}
    W_{\pm 1} = 1/2 \pm (1/2 - \eta) U(x_\mu,t) \label{Wpm}
  \end{eqnarray}
where the constant $\eta$ is a parameter of the model.
\item{The boid moves: $x_\mu$ is updated to $x_\mu + v_\mu$.}
\end{enumerate}

We take the neighbourhood of site $i$ to be the sites $i-1$, $i$ and $i+1$.
In the model of \cite{EOL} $U$ is given by the local
majority velocity which we now define. Let
the number of boids on a site with velocity $v$ be $n_v(x, t)$, the site density $\rho(x, t)=\sum_v n_v(x , t)$ and site
momentum $\phi(x , t)=\sum_v v n_v(x , t)$.  The majority velocity of
neighbouring boids is equivalent to the sign of the 
neighbourhood  (3 site) averaged momentum ${\widehat \phi(x, t)}$
given by
\begin{equation}
  {\widehat \phi(x,t)} = \frac{1}{3}\sum_{i=-1}^1 \phi(x+ia,t)\;.
\end{equation}
In what follows such a neighbourhood average of a quantity $X$ is
represented by
the hat symbol ${\widehat X}$. The preferred direction $U(x,t)$
is then  determined  by
\begin{equation}
  U(x , t) = \left\{\begin{array}{ll} 1,& \mbox{ if }
  {\widehat \phi(x,t)} > 0  \\ -1,& \mbox{ if } {\widehat
  \phi(x,t)} < 0  \\ 0,& \mbox{ if } {\widehat \phi(x,t)} = 0
  \end{array}\right.\label{originalbias}
\end{equation}
Thus in the model of \cite{EOL} the preferred direction is given
deterministically by the local momentum and the model thus includes
alignment.  Following a determination of $U$, $\eta$ is the
probability that the velocity is not updated to this value (and
instead to $-U$).  This probability is  independent of
the dynamical variables.

In this paper we generalise the model of \cite{EOL} at step (ii) above
to include the effects of centring and separation as well as
alignment---this will be discussed in section~\ref{CAA}.  We also
generalise to consider the case where $U$ is itself a stochastic
function of local variables: a boid may determine its preferred
direction from the local variables as $\pm 1$ or $0$ with some
probabilities (as will be motivated below).  However, since $U$
appears linearly in (\ref{Wpm}), averaging over its possible values
amounts to replacing $U$ by its expectation value ${\bar U}$ in
(\ref{Wpm}):
\begin{eqnarray}
 W_{\pm 1} = \frac{1}{2} \left[1 \pm G(x_\mu,t) \right] \label{Wpm2}
\end{eqnarray}
where
\begin{equation}
  G(x,t)= (1 - 2 \eta){\bar U}\;.
\label{G}
\end{equation}
Thus the important quantity is
${\bar U}(x,t)$,
the expectation value of the preferred direction
which should be a function of  the local density
and velocity, for example.
The expectation value
of the  updated velocity is given by $G(x,t)$.

In the next subsection we propose an explicit form for ${\bar U}$ and
hence $G(x,t)$. First it is useful to review the form of (\ref{G}).
The quantity $\mbox{sgin}(G)$ is the velocity to which a boid is updated in the
absence of noise, whereas $|G|$ is the certainty with which this
outcome is attained.  The uncertainty in the outcome comes from two
sources, the first is due to random errors, controlled by
the parameter $\eta$, where the boid moves against its preferred direction $U$.  
The second is from the stochastic nature of
$U$ itself, which as we argue in section~\ref{JOFFVCA} comes from the uncertainty
with which a boid perceives the local flock properties and therefore determines
its preferred direction.

\subsection{Alignment}
\label{AF}

In many models \cite{EOL,VCBCS,CBV,TT} it has been shown that
alignment alone is sufficient to produce complicated behaviour
recognisable as flocking.
Alignment is effected
in our dynamics through $G$. 
In this work we consider $G$ to be of the form 
\begin{eqnarray}
 G &=& (1 -2 \eta)\frac{\tanh(\beta V(x,t))}{\tanh(\beta)} \label{abias}
\end{eqnarray}
where
\begin{equation}
  {V(x,t)} = \frac{\widehat \phi(x,t)}{\widehat
  \rho(x,t)} 
\end{equation}
is the {\em neighbourhood average boid velocity}.
The parameter 
 $\beta >0 $
controls how nonlinear  $G(x,t)$ is as the function of
${V(x,t)}$.
In the limit where
$\beta\rightarrow\infty$ the majority rule case (\ref{originalbias}) is in effect.
In the limit $\beta\rightarrow0$ we obtain
a linear function.
The form (\ref{abias}) addresses several deficiencies in (\ref{originalbias}).
Firstly (\ref{abias}) is analytic, whereas (\ref{originalbias}) is non-analytic.
Secondly with (\ref{abias}) boids become sensitive
to increasing majority size, as might  be expected in physical systems.
In the following section we will see how this second intuitive feature can arise out of errors in local flock perception.

\subsection{Justification of form of $G$  via sampling argument}
\label{JOFFVCA}
We now turn to a justification of the form (\ref{abias}) from
microscopic considerations.  We present an argument amounting to a
simple algorithm for the determination of $G$ based on sampling of
neighbours.  In the model of \cite{EOL} boids determine $U$ with
perfect knowledge of all neighbouring boids (and perfect ignorance of
other boids); hence a local majority is determined with certainty.
Consider instead the situation where a boid can only make $M$
observations of its neighbours, each observation is of one boid
selected randomly from neighbouring boids with replacement. (Also note
that from the definition of neighbourhood a boid acts as a neighbour to
itself.)  The Majority rule  applied to the sample of $M$ boids then
requires a fixed observation time $O(M)$
regardless of local density. This is in contrast to the strict
majority rule algorithm which requires information about  all neighbours
(\ref{originalbias}), there being  no restriction on the
potential number of neighbours.

Consider a sample of $M$ neighbouring boids from the group of $3 {\widehat \rho(x , t)}$
neighbours, containing $3 {\widehat n_1(x,t)}$ rightward and
$3 {\widehat n_{-1}(x,t)}$ leftward moving boids.  With binomial
probability $P(k)$, $k$ will be selected from the rightward and $M-k$ from the
leftward moving groups:
\begin{eqnarray}
  P(k) &=& {M \choose k} \left(\frac{{\widehat n_1}}{{\widehat \rho}}\right)^k 
\left(\frac{{\widehat n_{-1}}}{{\widehat \rho}}\right)^{M-k} \label{Pk}\;.
\end{eqnarray}
The preferred direction  is determined from the sample
as  $U_M =\mbox{sign}(2k-M)$. The expected updated velocity $G_M$ is,
taking  $M$ is odd,  
\begin{eqnarray}
G_M &=& (1-2\eta) \overline{U}_M= (1-2\eta)\sum_{k=0}^{\frac{M-1}{2}}
\left[ P(M-k) - P(k)\right]\;. \label{UbarM1}
\end{eqnarray}
Using the substitution 
\begin{equation}
{\widehat n_j(x , t)} = {\widehat \rho(x ,
  t)}\frac{1 + j {V(x , t)}}{2} \; , \nonumber
\end{equation}
$G_M$ can be expanded as an  odd polynomial in $V$:
\begin{eqnarray}
  G_M &=& (1-2\eta)\sum_{k=0}^{\frac{M-1}{2}}\sum_{i=0}^{k}\sum_{j=0}^{\frac{M-1}{2}-k} \frac{{V}}{2^{M-1}} {M \choose k} {k \choose i} {M-2 k \choose 2 j + 1} (-1)^i {V}^{2(i + j)}\;. \label{UbarM}
\end{eqnarray}

A comparison of the two forms of $G$ (\ref{UbarM}) and (\ref{abias}) shows a qualitative correspondence between increasing $M$ and $\beta$.
For example, we have from (\ref{UbarM})
\begin{eqnarray}
  {G_1(x , t)} &=& (1-2\eta){V(x,t)} \nonumber\\
  {G_3(x , t)} &=& (1-2\eta)\frac{V(x , t)}{2}(3-{V(x , t)}^2)  \nonumber \\
  {G_5(x , t)} &=& (1-2\eta)\frac{V(x , t)}{8}(15 - 10 {V(x , t)}^2 + 3{V(x , t)}^4)\nonumber \;,
\end{eqnarray}
whereas  expanding (\ref{abias}) in powers of $V$ yields
\begin{equation}
  {G(x , t)} = (1-2\eta)\frac{\beta {V(x , t)}}{\tanh(\beta)}(1 - \frac{\beta^2}{3}{V(x , t)}^2 + \frac{2 \beta^4}{15}{V(x , t)}^4+O({V(x , t)}^6)) \nonumber \;.
\end{equation}
In  the two limiting cases $G_{M\rightarrow\infty}=\lim_{\beta\rightarrow\infty}G(x , t)$ and $G_{M=1}=\lim_{\beta\rightarrow 0}G(x , t)$ 
the correspondence is exact.

For intermediate values of $M$ it is possible to define an approximate
correspondence through a function  $\beta(M)$ using one of  several schemes. Anticipating our
mean field treatment below (section \ref{MF}) we choose
to match the gradient of
$G$ with respect to $v$ at the origin:
\begin{eqnarray}
    \beta(M) &=& \frac{1}{2^{M-1}}\sum_{k=0}^{\frac{M-1}{2}}{M \choose
    k}(M -2 k) = \frac{M}{2^{M-1}} {M-1 \choose \frac{M-1}{2}} \label{betaM} 
\end{eqnarray}
This yields, for example, $\beta(M=3 \mbox{ or } 4) = 1.5$
and $\beta(M=25 \mbox{ or } 26) \approx 4$.

To summarise we have argued that the form (\ref{abias}), parametrised by $\beta$, for the expected
updated velocity qualitatively corresponds to (but has a simpler form than) the result
of taking a sample of $M$ neighbours. The correspondence $\beta(M)$ is quantified in
(\ref{betaM}).

\subsection{Centring and Separation}
\label{CAA}

Centring, the tendency to move towards the local centroid
$\omega(x,t)$, defined as
\begin{equation}
  \omega(x,t) = \frac{\sum_{i=-1}^1 i \rho(x + i, t)}{\sum_{i=-1}^1 \rho(x + i, t)} \nonumber \;,
\end{equation}
 can be introduced through $G$ in an
analogous way to alignment. If a relative importance of $\kappa$ is
assigned to the effect of centring over
alignment in any observation then (\ref{abias}) can be
generalised  to
\begin{eqnarray}
  G(x , t) &=& (1 -2 \eta)\frac{\tanh\left( \beta \left( (1-\kappa)V(x,t) + \kappa \omega(x , t)\right)\right) }{\tanh\beta} \label{acbias}\;.
\end{eqnarray}

The desire for separation, typically through hard core repulsion, is
 very restrictive in one dimension.  If instead we consider separation
 as some cost associated with moving through a dense region then it
 amounts to a tendency to move away from the centroid (opposite to
 centring).  In order to include these effects together we introduce a
 capacity, $C$, which is the capacity at which the relative strengths of these two effects
 neutralise.  When the local density ${\widehat \rho(x , t)}$ is
 greater than $C$ the desire for separation exceeds the desire
 to centre, and vice versa.  Thus, our
most general form for $G$ is chosen as
\begin{equation}
  G(x , t) = \frac{(1-2\eta)}{\tanh\beta}
\tanh\left\{ \beta \left[ (1-\kappa)V(x,t) + \kappa \frac{C-{\widehat \rho(x , t)}}{C+{\widehat \rho(x , t)}} \omega(x , t) \right] \right\}  \label{acabias}\;.
\end{equation}
Note that (\ref{acbias}) is the limit $C\to \infty$ of (\ref{acabias}).

At this stage it is useful to summarise the parameters of the model with our choice of $G$:
the capacity $C$ is the density where the separation tendency
surpasses the centring tendency; $\kappa$ determines the relative
strengths of spatial density (centring and separation)
and alignment effects. Thus
$\kappa$ and $C$ determine the relative importance of Reynolds'
three flocking criteria.
The parameters $\eta$ and $\beta$ may
both be considered as sources of noise acting through $G$: $\beta$ introduces sensitivity to different field strengths
while $\eta$ introduces uncorrelated errors. 
With reference to section \ref{JOFFVCA} it is useful to consider the noise from $\beta$ as arising out of
stochastically sampling the  local flock to determine the boid's preferred direction
whereas $\eta$ encompasses additional error sources  once the preferred direction has been selected.

\section{Computer Simulations}
\label{CS}
\subsection{Observed Regimes}
\label{PAS}

\begin{figure}[hbtp]
  
    \centerline{
      \hbox{ 
	\epsfxsize 1\linewidth
	\epsffile{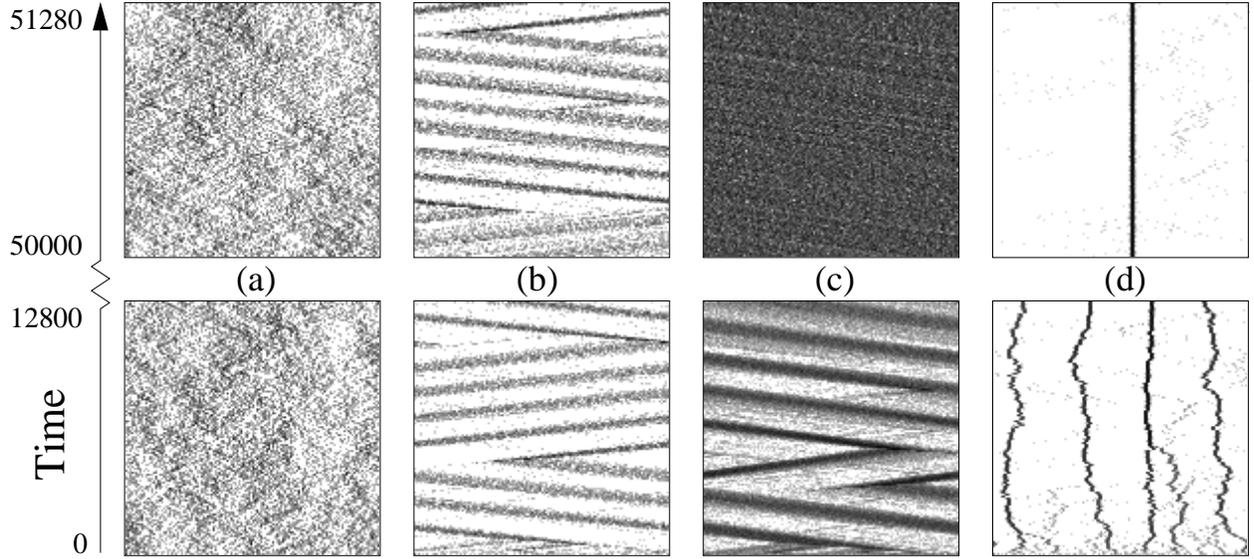}
	}
    }
  
  \vspace{0.2cm}
  \caption{\label{fig1} Figures show the development of systems in
  space time plots from random initial density and velocity
  distributions on systems with $L=128$ sites. The greyscale measures
  density $\rho(x,t)$ on a logarithmic scale. In this and following
  figures parameter sets are indicated [thus] (a) [$\eta=0.2, \beta=1$,
  $\kappa = 0.5$, $\mean{\rho}=1$, $C\rightarrow \infty$] A disordered
  system showing no sustained correlations in density or
  velocity. (b)[$\eta=0.02, \beta=8$, $\kappa = 0.5$, $\mean{\rho}=1$,
  $C\rightarrow \infty$] An alternating flock exhibiting a non-zero
  velocity between regular reversals. (c)[$\eta=0.125,
  \beta\rightarrow\infty$, $\kappa = 0$, $\mean{\rho}=8$] A
  homogeneous flock having homogeneous density and fixed global
  velocity.  emerging from transient alternations.  (d)[$\eta=0.02,
  \beta=8$, $\kappa = 0.75$, $\mean{\rho}=1, C\rightarrow \infty$] A
  dipole system consisting of several dipoles separated by low density
  homogeneous domains undergoes a slow coarsening process towards a
  single large dipole. (For the definition of a dipole see text.)}
\end{figure}

Numerical simulations were undertaken for a variety of noise levels
($\beta$ and $\eta$), and flocking parameters ($\kappa$ and $C$). Run
time constraints limited our investigations to systems of size upto
$1024$ sites and times upto $10^7$ timesteps, where one timestep is
sufficient time for each boid to be updated once on average ($N$
updates).  We found four characteristic and robust behaviours
which we now discuss, alongside a myriad of
intermediate  behaviours.

The first characteristic behaviour (figure \ref{fig1}a) is
the {\em disordered state}. This state has homogeneous density and a global
velocity of zero, global velocity being the mean velocity of a boid in
the system 
\begin{equation}
\mean{v} =\frac{\mean{\phi(t)}}{\mean{\rho}}\;.
\end{equation}
This
state persists at noisy parameter sets (high $\eta$, low $\beta$)
especially where Reynolds' effects exist at equivalent strengths or
where separation dominates ($\kappa\gtrsim 0.5$ and
$C\lesssim\mean{\rho}$).

The {\it alternating flock} is the second regime (figure \ref{fig1}b).
This system undergoes a repeating pattern of cohesive traversals
interspersed by rapid reversals, similar to that of \cite{EOL}.

The {\it homogeneous flock} as shown in figure \ref{fig1}c is a
flocking state which complements the alternating flock.  The
homogeneous flock has a homogeneous density and fixed non-zero global  velocity,
$\mean{v}$. 
This regime is found to be unstable where noise is high or where global density is low,
and is resilient only to small levels  of centring or separation.
The homogeneous flock is often established
following an alternating transient as shown in the lower part of
figure \ref{fig1}c. 
The homogeneous flock can also be observed in the model
of \cite{EOL} but at larger densities, or smaller system sizes than those considered there

The final regime observed is a {\it dipole state} (figure
\ref{fig1}d), which exists in systems where centring is the dominating
Reynolds' effect.  A dipole is a dense, localised structure. As we
shall discuss in detail (in section \ref{DR}) a dipole has a well
defined centre with $G$ pointing inwards trapping boids, and a
sharply decaying density profile at its edges.  A system of one or
many dipoles can exist on the array, either as direct neighbours or
separated by depopulated domains.  As can be seen in (figure
\ref{fig1}d) a slow coarsening process is observed in a system of many
dipoles, whereby small dipoles are eliminated and the boids are
absorbed by larger dipoles.  The dipole states exist for large values
of $\kappa$ and $C$ i.e. relatively weak alignment and separation.

\begin{figure}[hbtp]
  
    \centerline{
      \hbox{ 
	\epsfxsize 1\linewidth
	\epsffile{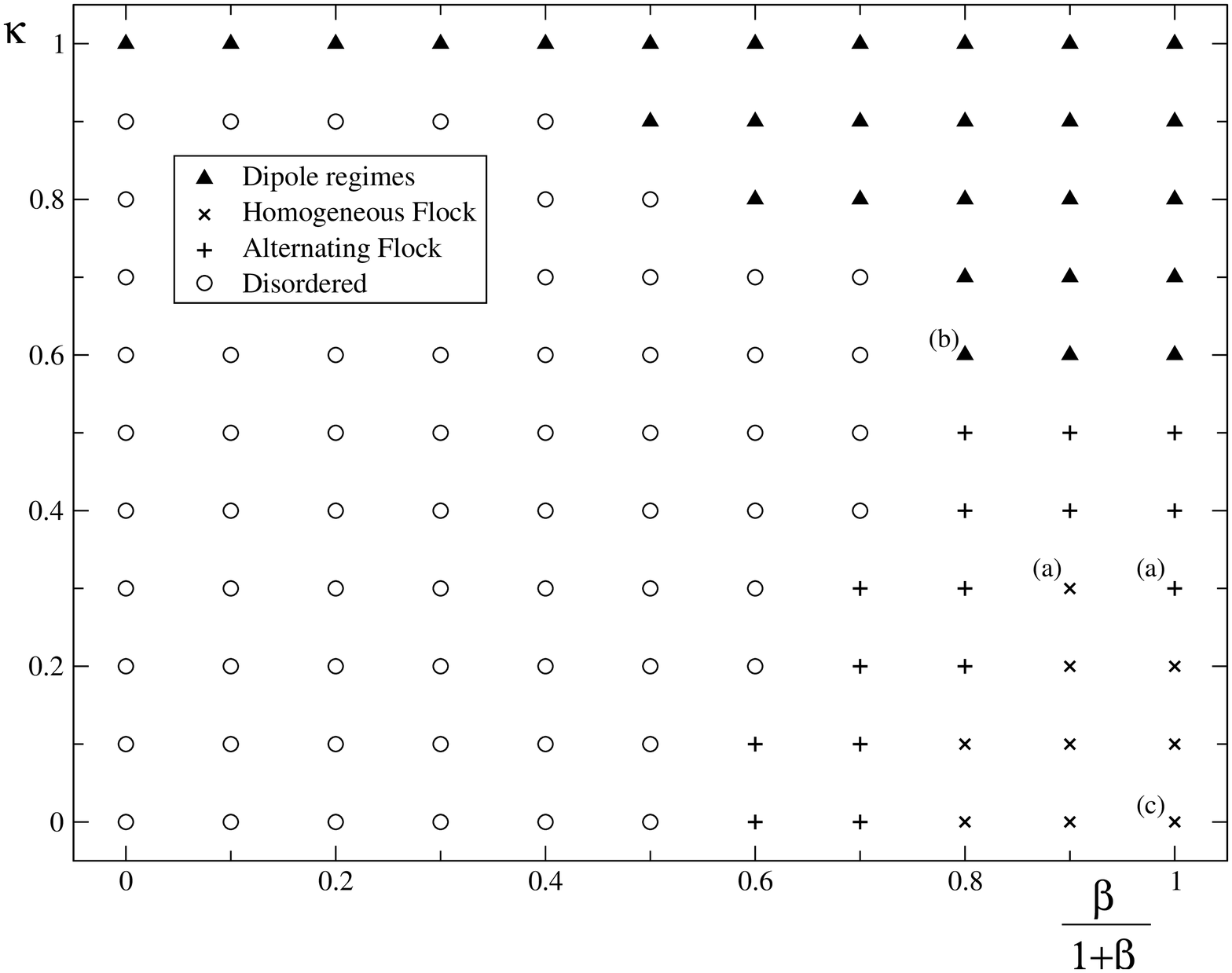}
	}
    }
  \vspace{0.2cm}
  \caption{\label{fig2} Diagram in the $\kappa$-$\beta/(1+\beta)$ plane
showing the regimes present in a
system with $L=128$, $\mean{\rho}=4$, $\eta=0.05$ with alignment and
centring effects only ($C\rightarrow\infty$). The dipoles,
alternating flock and homogeneous flock all occupy characteristic
regions. Intermediate  states at the boundaries of these
regions   are characterised by mixed behaviours:
(a) Flocks behave for sustained periods as both homogeneous flocks
  and alternating flocks, (b) dipoles begin to traverse large
  distances in single motions. The model of \cite{EOL} is case (c), a
  homogeneous flock for this parameter set and system size.}
\end{figure}

These four regimes all have readily characterised behaviours and
occupy characteristic regions of parameter space as shown in figure
\ref{fig2} which describes the case $C\to \infty$ (absence of separation).
Figure \ref{fig2} may be thought of (and we shall refer to it)
as a phase diagram. However, the different regimes are not truly
``phases'' as the boundaries between them depend on system size and, as we shall
see below, the  homogeneous flock regime actually
disappears in the limit of large system size.
In the
following subsection we shall discuss in more detail
the flocking and dipole regimes.

\subsection{Flocking regimes}
\label{FR}
The alternating flock was originally studied in \cite{EOL} where only
alignment was present. Here we find that it persists in systems with
all three Reynolds' effects present provided alignment dominates
($\kappa \lesssim 0.5$) and $\beta$ remains finite.  In its traversing
stage a single dense group of boids (the flock), confined within a
non-spanning section of the lattice, slowly flows 
across the system at a consistent non-zero velocity and diffuses.  At a later time
the {\it reversal mechanism} begins: this is initiated by a
fluctuation near the front of the flock causing a local reversal in
momentum.  The fluctuation moves  against the direction of the  flock and grows in
density and momentum until it passes entirely through the flock
inverting its velocity.  At this time the traversing behaviour of a flock with
opposite velocity is established.  Initiation of reversals can occur
either internally or by collision with boids external to the
flock. While not all large momentum fluctuations initiate a full
reversal, they occur often enough
to reverse the flock before it
spreads out to occupy the whole lattice through  diffusion.
The reversals occur stochastically with no fixed period but there is a well
defined mean time between reversals.

\begin{figure}[hbtp]
    \centerline{
      \hbox{ 
	\epsfxsize 0.5\linewidth
	\epsffile{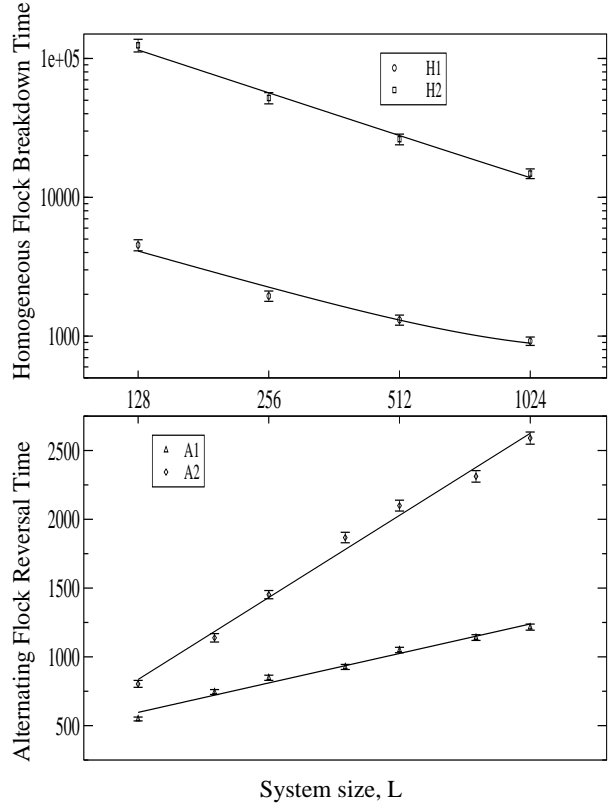}
      }
    }
    \vspace{0.2cm}
  \caption{\label{fig3n} 
a) Breakdown times  for  homogeneous flocks
and b) reversal times for alternating flocks
vs $\ln L$ where $L$ is the system size.
Data is averaged over  100 and 1000 breakdowns and reversals respectively.
(a) H1: $[\eta = 0.0413, \beta=1, \kappa=0, \mean{\rho}=10]$, H2:
$[\eta = 0.125, \beta = \infty, \kappa =0, \mean{\rho}=8]$.
The straightline is a least squares fit to   $T = \frac{L}{4}+A
  L^{-B}$ (see text). 
The exponent is   $B=1.03$ for H2,
  and $0.89$ for H1. (b) 
A1: $[\eta= 0.005, \beta=2, \kappa=0, \mean{\rho}=1]$, A2:
$[\eta = 0.02, \beta = 8, \kappa =0.25, \mean{\rho}=1, C= \infty]$.
The reversal times in alternating flocks are
logarithmic and the straight lines are least squares fits to the data.}
\end{figure}

In the alternating flock instabilities occur at the front edge of the
flock.  In \cite{EOL} it was established that the reversal timescale
for alternating flocks is $O(\ln N)$ where $N$ is the number of boids.
The argument is that this is the timescale on which the front edge
becomes susceptible to {\it relatively} large momentum fluctuations,
due to its low density.  In our model there is a new form for $G$, but
logarithmic timescales appear to persist for the alternating regime as
shown in figure \ref{fig3n}b.  

\begin{figure}[hbtp]
    \centerline{
      \hbox{ 
	\epsfxsize 0.7\linewidth
	\epsffile{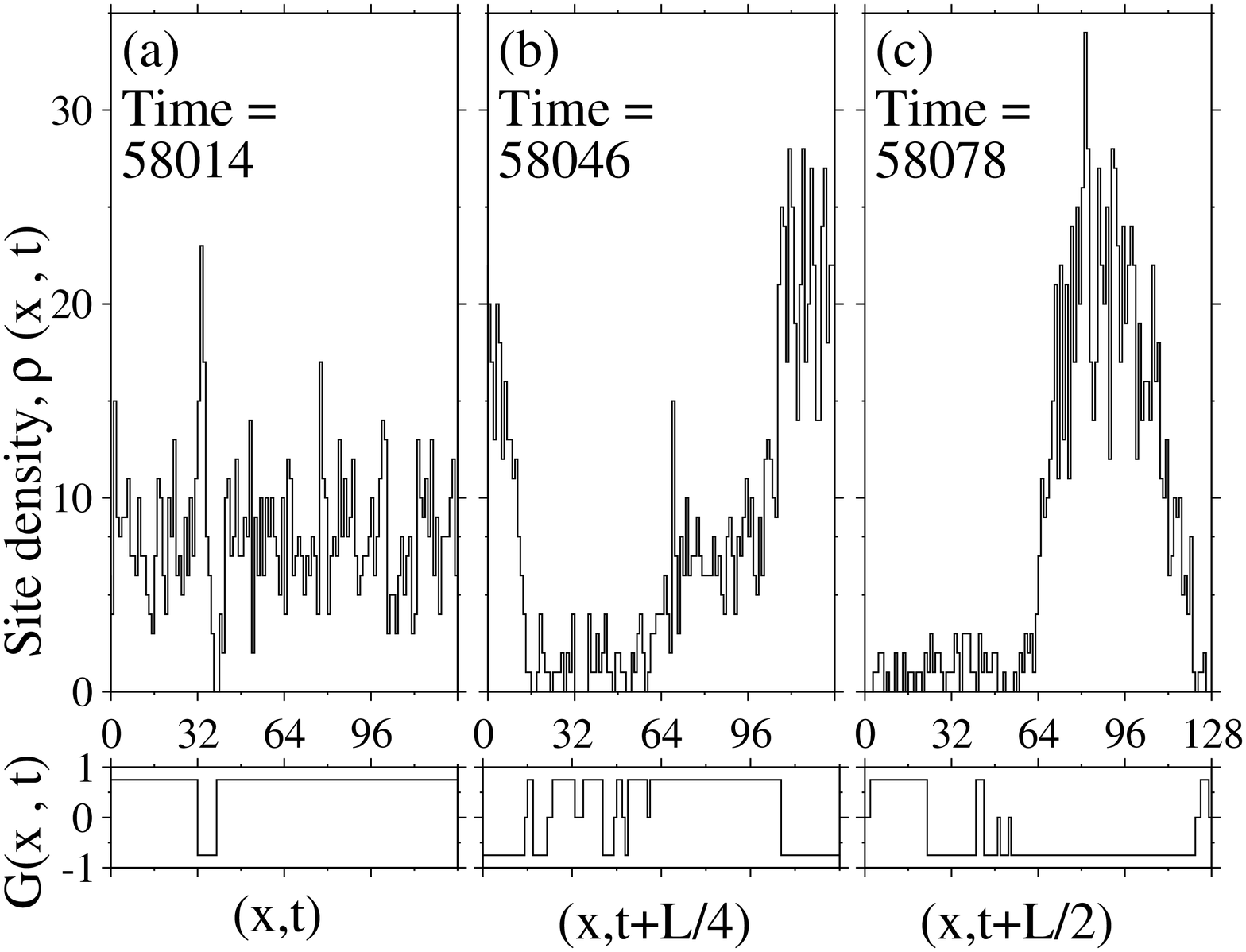}
      }
    }
    \vspace{0.01cm}
    \centerline{
      \hbox{ 
	\epsfxsize 0.7\linewidth
	\epsffile{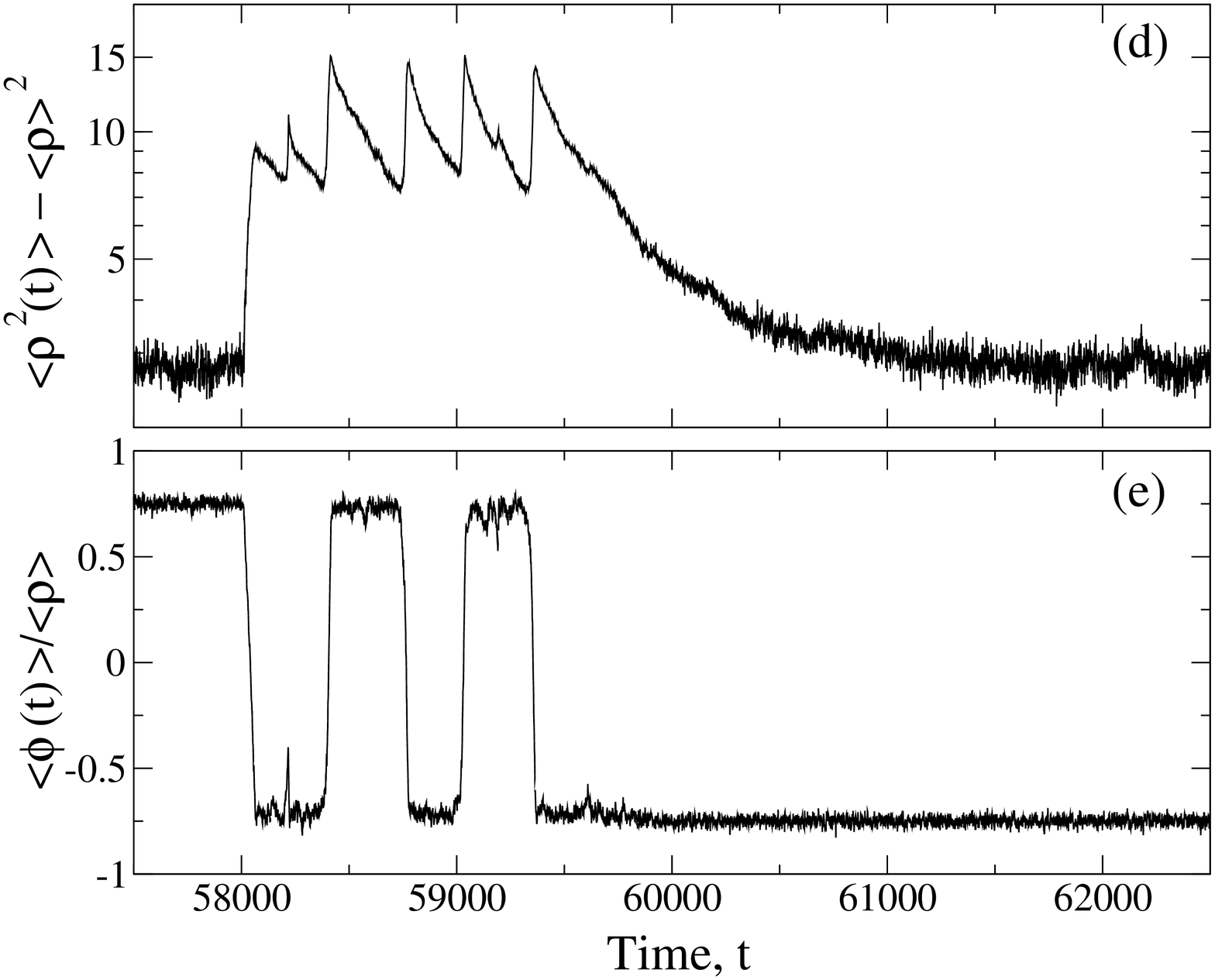}
      }
    }
    \vspace{0.1cm}
\caption{
\label{fig4n} 
The Homogeneous flock of \ref{fig1}c undergoes a rapid change of behaviour.
\newline
(a,b,c) Density and $G$ are plotted as a function of position at fixed
times.  a) $t=$58014: values are consistent with a homogeneous flock
throughout except near site $32$ where a localised momentum
fluctuation exists.  (b)$t=$58046: The interface between the two
opposing domains created moves leftwards and the fluctuation gathers
momentum.  (c)$t=$ 58078: Within a short space of time the entire
flock is reversed, and the density and $G$ profiles are characteristic
of an alternating flock.  Following this event an alternating flock is
temporarily established, before an homogeneous flock reemerges.
\newline
(d,e) A measure of the spatial distribution of boids and the global
velocity plotted against time from before the breakdown, through the
temporary alternating flock until the emergence of a homogeneous flock.}
\end{figure}

As was  shown in  figure \ref{fig1}c, from  many initial  conditions a
homogeneous  flock  develops   after  an  initial  transient alternating  flock.
However many  homogeneous flocks  are temporarily unstable  towards an
alternating flock  profile even after  a long time in  the homogeneous
regime.  Such  a process  occurs for the  homogeneous flock  of figure
\ref{fig1}c after 58000 timesteps.   In the breakdown process a fluctuation in
momentum, initially confined to  a few  sites, creates  a disturbance
which moves  through  the flock in a manner  similar to the reversal
mechanism  in  alternating flocks.  Figures  \ref{fig4n}a,b,c show  the
progression  of the  fluctuation through  the flock  to a  point where
global velocity is completely inverted.  Before the flock returns to a
homogeneous flocking  regime there are several  further reversals. The
localisation of  the boids and  high reversal rate during  this period
characterise    a transient    alternating    flock    state    (see    figures
\ref{fig4n}d,e).

It appears that the temporary failure of the homogeneous flock arises out
of a large local fluctuation in momentum.  During flock reversal the
initial fluctuation acquires boids from the oncoming flock and
continues to grow in momentum and stability.  Although it appears that
the size and type of disturbance required might vary for different
systems, the rate at which such disturbances occur over the  system should be
proportional to the number of nucleation sites.  Within the
homogeneous flock, there is a homogeneous density so the number of
nucleation sites is proportional to  $L$, thus we expect
the breakdown time for the flock, $T_B$ to behave as $1/L$.  In figure \ref{fig3n}a we take the mean time
a  homogeneous flock of positive momentum first reverses and attains a
negative global velocity as the flock breakdown time.
We fitted  the data by
 $T_B\approx \frac{L}{4}+AL^{-B}$, where we expect $B$ to be approximately
$1$. The constant $\frac{L}{4}$ is our approximation to the 
time for  the fluctuation to pass  through half the flock. Such a prediction
appears to closely match  computer simulations as illustrated in Figure~
\ref{fig3n}a.

The  previous paragraph implies  that for large systems
the homogeneous flock will always become unstable due  to nucleation
of  momentum fluctuations.
To quantify this, consider how 
the breakdown of the homogeneous flock
to an alternating regime is  initiated by a sequence of
successive microscopic flips against the preferred direction  sufficient to invert the momentum. The
number of flips required for this to occur will be proportional to the
local density, $\mean{\rho}$, and number of sites over which $G$ is
determined (range, $r=3$ in our simulations). Given that a flip occurs with
probability $P_F =\frac{1-|G|}{2}$, flock breakdown will be
initiated with probability $\sim L P_F^{k r \mean{\rho}}$ in any
timestep. Therefore one expects the typical time $\tau $ for the flock to flip
to be  $\tau \sim L^{-1} P_F^{-k r \mean{\rho}}$. This   implies
that for the homogeneous flock to be stable over times
$\tau \sim L^{\alpha}$ where $\alpha >0$, a density    $\mean{\rho} \sim \ln L$ would be required.
We conclude that for
finite range and density no homogeneous flocks will be stable in the
thermodynamic limit ($L\rightarrow\infty$). However, as we have seen
in Figure \ref{fig2} on intermediate size systems   there are well
defined regions of parameter space where homogeneous flocks do exist.

Finally we examine the impact of varying the strength of centring and
separation on flocking states.  The alternating flock persists even
when centring is a relatively strong effect, whereas the homogeneous
flock typically becomes unstable towards an alternating flock under such conditions.  Thus
in the phase diagram in figure \ref{fig2}, for large enough $\beta$,
increasing $\kappa$ leads to transitions from the homogeneous to the
alternating flock and thence to the dipole state.  The phase diagram
with the inclusion of separation (finite $C$) is qualitatively
similar.  Perhaps surprisingly, decreasing $C$ does not substantially
affect the alternating flock.

\subsection{Dipole Regimes}
\label{DR}

\begin{figure}[hbtp]
  \centerline{
    \hbox{ 
      \epsfxsize 1\linewidth
      \epsffile{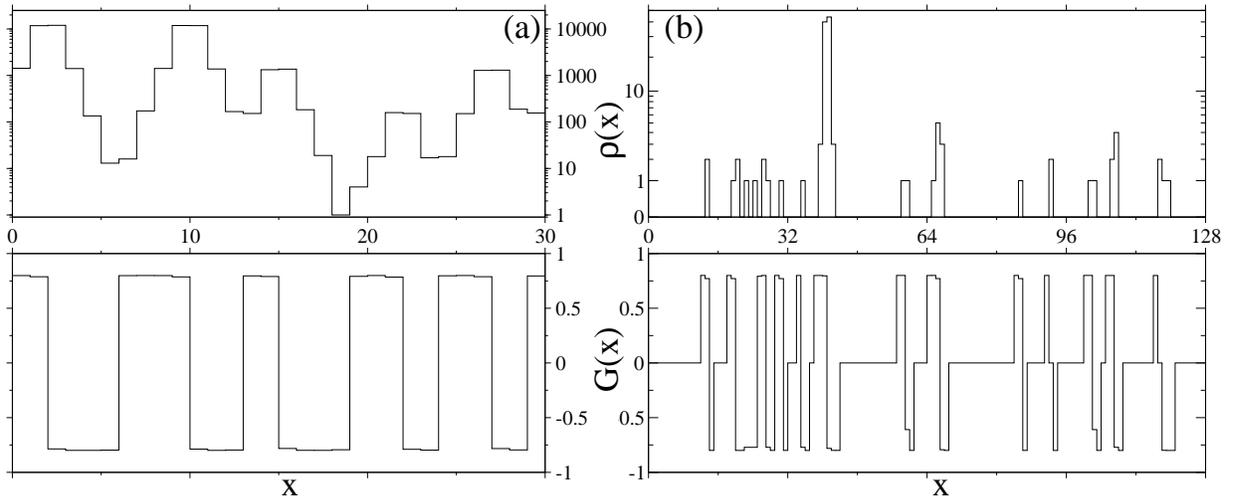}
    }
  }
  \vspace{0.2cm}
  \caption{\label{fig5} [$\eta=0.1, \beta=4$, $\kappa = 0.75$,
$C\rightarrow\infty$]. Density Profiles (upper panels) and 
profiles of the expected updated velocity, $G$, (lower panels) after 50000 timesteps. (a) At high
density $\mean{\rho}=2000$ domains,
with  exponentially decaying densities and
fixed velocity, bounded  by discontinuities in $G$ are clear. (b) At density
$\mean{\rho}=1$ dipoles occupy a small portion of the lattice, but
contain the majority of boids. There are large portions devoid of boids with $G=0$.}
\end{figure}

The dipole regime is characterised by single or multiple localised
structures.  These may be stationary or slowly moving. The density is
maximal at the centre of the dipole and the density profile is
symmetric about the centre.  However, to the left of the centre $G$ is
 positive but to the right of the centre $G$ is
negative. Thus the structure traps particles in such a way that the
net flux of boids across the centre of the dipole is zero.  

In figure
\ref{fig5} we illustrate two systems, one at high density and the
other at densities typical of our other simulations (approximately 1).  In
the case of high density it can be seen that the system is dominated
by adjacent large dipoles.  Each of these dipoles has an exponentially
decaying density to a boundary with another dipole, and a steady value for $G$ trapping boids. 
In this system there is slow redistribution of boids
across boundaries towards larger dipoles.  In the second system
density is lower and the exponential density profiles are not obvious;
instead there is one large dipole and several smaller dipoles. The  dipoles span a very small
fraction of space and are not in direct contact with others. The gaps between the dipoles
are essentially homogeneous regions of very
low density but small dipoles (of size 2--5 boids in the figure)
regularly form and evaporate away. Occasionally
a small dipole arising  out of the low density region
can grow to supersede larger dipoles.

\begin{figure}[hbtp]
    \centerline{
      \hbox{ 
	\epsfxsize 0.8\linewidth
	\epsffile{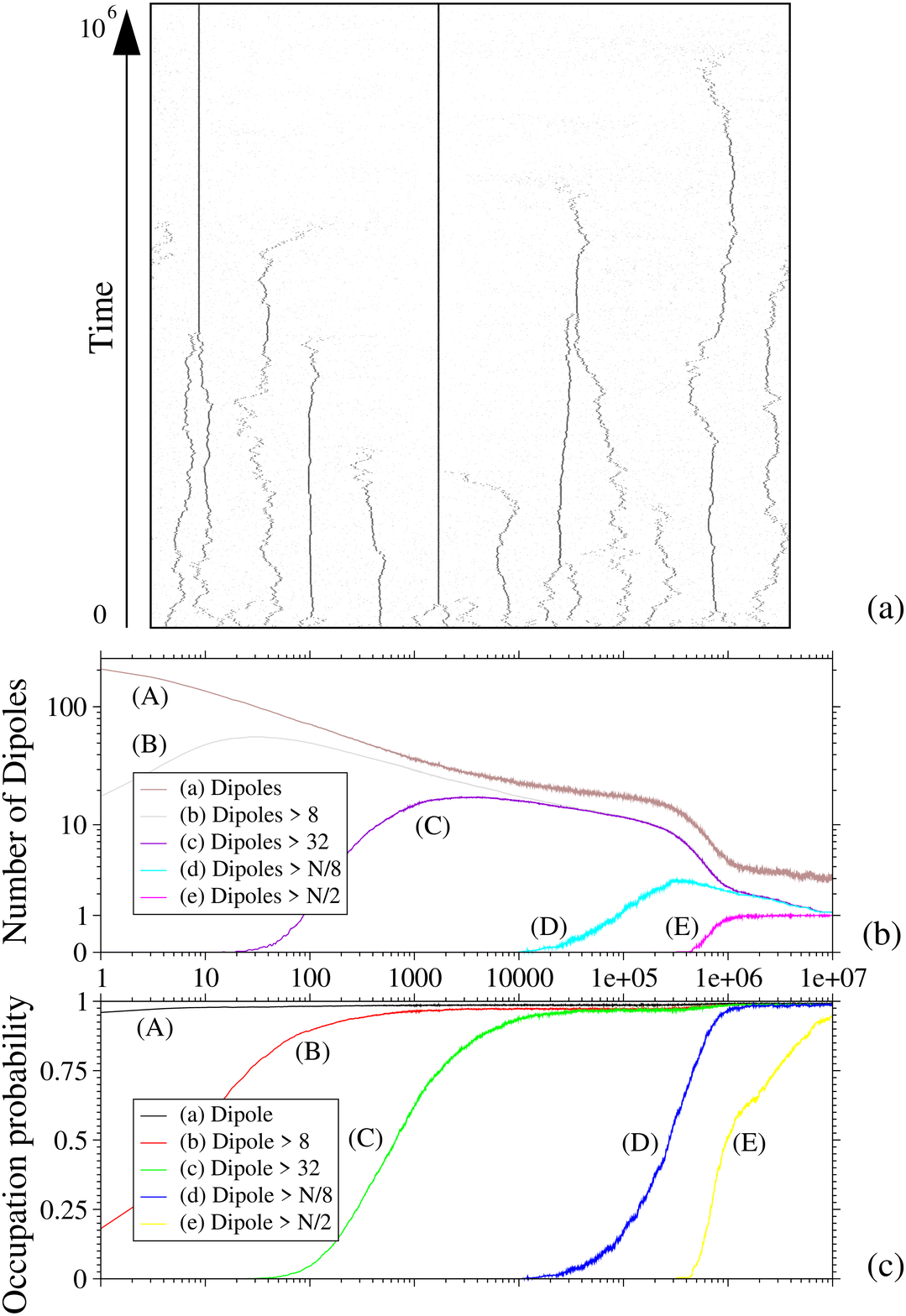}
      }
    }
    \vspace{0.2cm}
  \caption{\label{fig6} 
System with same parameter set as  figure \ref{fig1}d  on a
larger system  of $L=1024$.
 (a) A typical space-time plot 
starting
from random initial conditions. After an initial transient period when the dipoles form,
an intricate
coarsening process takes place   wherein small  dipole  eventually dissolve
and large dipoles grow.
  (b,c) Statistical data on the  remaining dipoles  as a function of time
averaged from $100$ runs similar to (a). (b) The number of dipoles
remaining  of size greater than the values indicated in the legend: 
Eventually the number of dipoles of intermediate size decreases to zero (curves B,C,D,E converge to 1);
After $10^7$ timesteps a single large dipole emerges alongside approximately $2$ small transient dipoles. 
(c) The probability that a boid is contained within a dipole greater than the sizes
indicated in the legend as a function of time:
Only a small number of boids ever exist outside dipoles (A) and after sufficient time the majority of boids are contained in a single large dipole (E).}
\end{figure}

In figure \ref{fig6} details of a typical coarsening process, whereby
small dipoles are eliminated and large dipoles grow, is illustrated.
Figure \ref{fig6} shows the development of a system, after an initial
transient period when many dipoles form, through an intricate
coarsening process.  Figure \ref{fig6}a shows the system coarsening
into two large dipoles and eventually (at a later time not included in
\ref{fig6}a) one large dipole emerges. This may be seen from figures
\ref{fig6}b,c which plot the statistics of the sizes of the dipoles
remaining in the system.  The regions in between the dipoles is a low
density domain containing independent boids which combine sporadically
to form small transient dipoles.  In the fully coarsened system there
is a relatively small probability of a boid being elsewhere than the
single dominant dipole---all dipoles of intermediate size disappear.

In the low density regime (as in \ref{fig5}b) the exchange of boids
between dipoles occurs through the homogeneous low density domains.
Dipoles may lose boids through stochastic fluctuations at their
boundaries.  If two adjacent dipoles have such noise-induced boid loss
at different rates, there will be a net flux across this domain
towards the more stable dipole.  Numerically we have observed that in
fact smaller dipoles have a higher loss rate than larger ones.
Therefore the rate of boid loss depends on the fine details of the
dipole profile which depend on the dipole size. (An assumption of a
pure exponential density profile for all dipoles would imply no
dependence of the rate of boid loss on dipole size.)  Other effects
might also be important, for example the greater mobility of smaller
dipoles.

Dipoles are created and sustained where centring is the dominant
effect. Within dipole systems increasing alignment (decreasing
$\kappa$) is most noticable in the resulting greater mobility or {\it
wandering}\cite{LR} of dipoles (and independent boids); a significant
process since it hastens the coarsening of the system, especially in
the initial stages. Wandering can occur in any size dipole but is most
prominent in smaller dipoles. Typically wandering is initiated by a
net flux of boids across the centre of the dipole causing the centre
to shift. This motion can be induced or enhanced by alignment, since
the matching of velocities in such a process favours the collective
motion.  With alignment sufficiently strong, this process can be 
self-reinforcing to the extent that a sequence of movements or sustained
translations (flocking) become possible.

\begin{figure}[hbtp]
    \centerline{
      \hbox{ 
	\epsfxsize 0.5\linewidth
	\epsffile{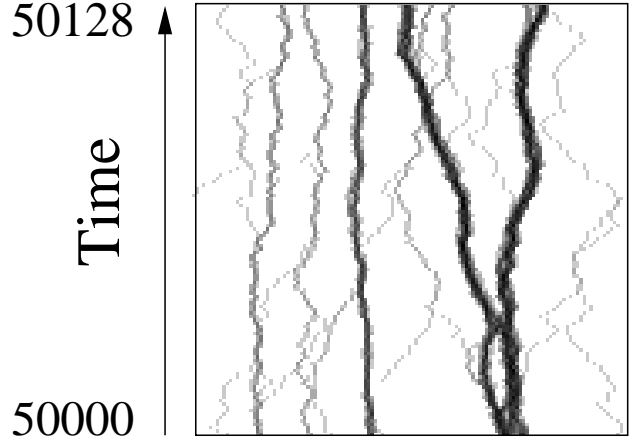}
      }
    }
    \vspace{0.2cm}
  \caption{\label{fig7} [$\eta=0.02, \beta=8$, $\kappa = 0.85$,
$\mean{\rho}=1, C=20$] This system contains two large  fixed-density
dipoles  alongside small dipoles and solitary boids. In small dipoles
centring is the dominant effect and wandering is limited, however
in the two  fixed-density dipoles large translations are observed.
See text for explanation of the term fixed-density dipole.
}
\end{figure}

Dipole systems are characterised by their central region of high
density. However, we  might expect such a profile
to be suppressed by separation. The inclusion of separation in dipole
systems leads to a dramatic alteration of behaviour: where alignment
is a small effect, and $\mean{\rho} < C < \infty$, a system of {\it fixed-density
dipoles} can be observed. Fixed-density dipoles have reduced density
at the centre, but the density is constant within an extended central region
and only near the boundaries does one see the
sharply  decaying density profile, thus the density profile has a flat top. 
Within fixed-density dipoles alignment plays a much more
important role than in comparably sized dipoles in the absence of separation, since at the centre
of fixed-density dipoles the separation and centring tendencies tend to
cancel allowing alignment to dominate. The relative stability of large
and small fixed-density dipoles is left as an open question. Fixed-density
dipoles can, for example, split from their centre and wander more
substantially. These processes are appreciable in figure \ref{fig7} where the mobility of
 the large fixed-density dipoles is second only to that of solitary boids.

\section{Mean Field Theory}
\label{MF}
In order to understand some of the regimes observed numerically above, we now
develop,  from a mean field treatment of the full
dynamics, a continuum model in which
the density $\rho$ and neighbourhood velocity $v$ are continuous functions.
In so doing we are able to identify a linearly stable
flocking solution comparable to the homogeneous flock and a piecewise
continuous solution comparable to the  dipole
regime. Furthermore  we show that these solutions exist in numerical
iterations of the mean-field equations.  

\subsection{Equations}
\label{MFTD}
The equations satisfied by the mean-field 
density, $\rho$, and mean-field neighbourhood velocity, $v$, are derived in appendix A
and read
\begin{eqnarray}
  \frac{\partial \rho}{\partial t} &=& - \frac{\partial}{\partial x} (\rho G) + \frac{1}{2}\frac{\partial^2}{\partial x^2} \rho \label{continuumrho}\\
  \frac{\partial v}{\partial t} &=& -v + G + \frac{1}{\rho}\left[v\frac{\partial}{\partial x} (\rho G)-\frac{\partial}{\partial x} \rho\right] \label{continuumv}
+ \frac{1}{2\rho}\left[- v\frac{\partial^2}{\partial x^2} \rho+\frac{\partial^2}{\partial x^2} (\rho G)\right]
\end{eqnarray} 
where
\begin{equation}
  G(v,\omega) = \frac{1 -2 \eta}{\tanh \beta}\tanh\left[ \beta \left( (1-\kappa)v + \kappa \omega\right)\right] \label{MFU}
\end{equation}
and
\begin{equation}
  \omega = \frac{2 }{3 \rho}\frac{\partial}{\partial x} \rho\;. \label{OmegaMF}
\end{equation}
Equation (\ref{continuumrho}) is rather easy
to understand: note that, since $G$ is the mean velocity of boids,
$\rho G$ is a mass current, thus the equation is a continuity
equation with a  current  and a diffusive term.
The equation  for the neighbourhood velocity $v$ (\ref{continuumv}), on the other hand is
more complicated.

\subsection{Steady State Solutions and Linear Stability}
\label{LS}
We now look for steady state solutions of the above equations, for
which we set the lhs of (\ref{continuumrho},\ref{continuumv}) to zero.
Further, we look for homogeneous solutions where spatial derivatives are zero.
The global density may be freely
chosen, in these solutions $\rho(x,t) = \mean{\rho}$, whereas
velocities are determined  by (\ref{continuumv}) as 
$v= G(v,0)$ which yields
\begin{eqnarray}
  v &=& (1-2\eta)\frac{ \tanh\left( \beta (1-\kappa) v \right)}{\tanh \beta} \;. \label{transcendental1}
\end{eqnarray}
Thus, there are either one or three solutions to the mean field
theory which are homogeneous in density.  

For all parameter sets one solution to this is $v=0$, the disordered
solution, whilst for certain values of $\beta$, $\kappa$ and $\eta$
two {\it flocking solutions} $v=v_\pm$ exist, corresponding to the
homogeneous flock.  The existence of the flocking solutions is
determined by the gradient in $G$ with respect to $v$ at the
origin. Since $G$ is a concave function of $|v|$ a flocking solution of
(\ref{transcendental1}) requires that $G'(0,0) \geq 1$. Thus the
existence of this  flocking solution is dependent on the
following criterion being satisfied:
\begin{eqnarray}
  \frac{\tanh \beta}{\beta} &\leq& (1-2\eta)(1-\kappa)   \label{solc}\; .
\end{eqnarray}
This is possible in systems with low $\eta$, but strong alignment (so that $\beta (1-\kappa)$ is large).

\subsection{Piecewise Dipole Solution}
\label{PDS}
We show in this section how it is possible to determine a further steady-state solution of
(\ref{continuumrho},\ref{continuumv}) where the density is an exponentially
increasing or decreasing  function of $x$ and $v$ is a positive or negative constant
respectively. 
Domains of these solutions may be pieced together
so that the density is continuous to form  dipole structures.

First, we assume a fixed value of $\omega$ implying by definition
(\ref{OmegaMF}) a density profile
\begin{eqnarray}
  \rho(x , t) &=&  A \exp\left( \frac{3 \omega x}{2}\right)\label{dipolesolutionrho}
\end{eqnarray}
where $A$ is a constant.
Next we propose a solution whereby within
any domain $G$ and hence velocity are fixed. 
Then, from (\ref{continuumrho}) we have
\begin{equation}
  0 = - G(w,v)\frac{3}{2}\omega \rho(x) + \frac{9}{8}\omega^2\rho(x)
\end{equation}
which implies
\begin{equation}
  G(\omega,v) = \frac{3\omega}{4} \label{Gomega}\;.
\end{equation}
The velocity $v$ must similarly be steady and (\ref{continuumv}) implies
\begin{eqnarray}
  0 &=&  - v + G(\omega,v)+ \frac{3}{2}\omega G(\omega,v)v
- \frac{3}{2}\omega - \frac{9}{8}\omega^2 v + \frac{9}{8}\omega^2 G(\omega,v) \nonumber
\end{eqnarray}
which after substitution of $G$ from (\ref{Gomega}) simplifies to
\begin{equation}
  v = \frac{3\omega}{4}\left[ 2 \left(\frac{3\omega}{4}\right)^2 - 1\right]\;.
\label{v(omega)}
\end{equation}

The condition required for a non-homogeneous solution is the consistency of (\ref{Gomega}) and (\ref{MFU}), eliminating $G$ we obtain
\begin{equation}
  \omega = \frac{4 (1 - 2 \eta)}{3 \tanh{\beta}}\tanh\left\{\beta \left[\left(\frac{7\kappa}{3}-1\right)\frac{3 \omega}{4} + 2(1-\kappa)\left(\frac{3 \omega}{4}\right)^3\right]\right\}\label{dipoleomega}\\
\end{equation}

Alongside the homogeneous density solutions, $\omega=0$, there are
potentially dipole solutions of (\ref{dipoleomega}) $\omega_\pm\neq0$.
Correspondingly $v$ takes values $v(\omega_\pm)$ via (\ref{v(omega)}).
One of these has rightward $G$ and exponentially increasing density,
the other a leftward $G$ and exponentially decreasing density.  A
dipole is a localised structure consisting of a left hand region of
exponentially increasing density and a right hand region of decreasing
density.  To satisfy periodic boundary conditions it is necessary that
the system consists of complementary domains.  Consider a boundary
between the dipole domains: the two exponentially decaying or
exponentially growing domains meet with equal density at such a point.
The density gradient and other qualities are locally antisymmetric
about the boundary; hence the net flux of boids will be zero and the
boundary stationary (this would not be true for a dipole domain in
contact with a homogeneous domain). Within each domain the presence of
a boundary is not felt, and so domains of differing width and height
can neighbour one another. Hence non-symmetric dipoles of differing
sizes can be neighbours.  Our solutions are therefore the set of
coupled non-overlapping dipole systems (domain pairs $i$), each
defined uniquely by the set of dipole centre positions, heights and
widths.

These solutions are comparable to the full stochastic system at very
high density (see figure \ref{fig5}a).  Note that the larger dipoles
in the sparser system (figure \ref{fig5}b) also approximate a pair of
domains with the same fixed $G$ and $\omega$.

In numerical iterations of the mean field equations  (see next subsection)
the dipole structures were rounded at the cusps which form  at domain boundaries.
One expects  this rounding and the physics at the cusps
to be described by higher order terms  in the lattice spacing $a$ which have been ignored (see Appendix A).

\subsection{Numerical Analysis}
\label{NA}
Numerical iteration of the dynamical mean field equations
(\ref{continuumrho},\ref{continuumv}) allows further analysis of these
states and comparison to full system dynamics. The iteration was carried out
by rediscretising  (\ref{continuumrho},\ref{continuumv}) onto a lattice. We found the flocking
($\omega=0,v_\pm$), non-flocking ($\omega=0,v=0$) and dipole solutions
($\omega_\pm,v(\omega_\pm)$) to be present, in addition to systems
comparable to the fixed-density dipole.  The alternating flock however
was never observed, nor were systems comparable to sparser dipole
regimes.

Where a flocking solution is predicted, and not made unstable by the
presence of centring (see section~\ref{stab}) a homogeneous flock with
the anticipated global velocity (\ref{transcendental1}) was always
observed to emerge from any initial conditions.

In dipole regimes we found that the system iterated to a dipole
system consisting of many dipoles.  However the mean
field dipole system does not exhibit a further coarsening process
comparable to the full stochastic model.  This is likely due to the
absence of noise driven fluctuations which drive the coarsening
process.  On the other hand, where the initial boundary between
domains does not fall perfectly on a lattice space boundary or centre
a very slow redistribution of boids does occur in the mean-field system, but apparently
at only an exponentially slow rate. This redistribution was not
observed to cause the disappearance of any dipoles, nor did it appear
to favour transfer to larger dipoles (is  the case in the full stochastic system).

\subsection{Linear Stability of Mean Field Solutions}
We have  developed a linear stability analysis of the disordered and
flocking solutions to the mean field equations.  This is done by
linearising (\ref{continuumrho}, \ref{continuumv}), computing the
evolution of the Fourier components then checking whether positive
(unstable) eigenvalues exist.  As the details are rather involved we
do not present them here but quote some results.  
In the case
$\kappa=0$, when the flocking solutions exist they are linearly stable
and the disordered solution is unstable; when the flocking solution
does not exist the disordered solution is stable. However, for
sufficiently large $\kappa$ an instability will occur in both flocking
and disordered solutions. This instability is expected to be with
respect to stable dipole solutions, although we did not explicitly
check the stability of dipole solutions.

Thus the equality in (\ref{solc}) defines a mean-field phase boundary
between the disordered and possible flocking states.  Comparing to
Figure~\ref{fig2} this curve is in approximate agreement to the
boundary between the disordered and alternating flock regimes.
However for other parameter sets the boundary is less accurately
predicted. Generally the boundary becomes increasingly more accurate as the
density increases.

As well as the disordered/flocking phase boundary,
the instability of the flocking solutions at large $\kappa$ is also consistent with
Figure~\ref{fig2} where for large $\kappa$ dipole solutions appear. 
The only discrepancy remains the absence of alternating
flocking solutions within the mean field theory: in regions where one
sees an alternating flock in the full stochastic system the mean field
theory has a homogeneous flock solution.

\subsection{Absence of Alternating Flock and Stability of Mean Field Solutions}
\label{stab}
As noted above the alternating flock was not expressed within the mean
field theory, even in the time dependent form (\ref{continuumrho}, \ref{continuumv}). 
 As we shall now discuss, the mean field theory is not
capable of describing the alternating flock which is driven by
fluctuations.  However the flocking solution of the mean field theory
does appear to correctly predict the velocity of the alternating flock
between reversals  i.e.  the solutions of (\ref{transcendental1})
proves to be in good agreement with transient results from the full stochastic
system.
We also  note that we found numerically that the mean field theory correctly
describes the evolution of the profile of an alternating flock as it
spreads between reversals.  However, whereas an alternating flock in
the full stochastic system will inevitably reverse due to some
fluctuation in momentum (as described section \ref{FR}) the mean-field
profile will keep evolving until the homogeneous flock is attained.
Similarly, due to the absence of fluctuations the homogeneous flock
will not reverse for any system size.

Clearly, to reproduce the alternating flock, a noise term should be
added to the mean field equations (\ref{continuumrho},
\ref{continuumv}) thus turning them into Langevin equations.  This
would be done, following the proposal of \cite{VCFH,CBV}, by adding a
noise term to the velocity equation (\ref{continuumv}).  The equations
(\ref{continuumrho}, \ref{continuumv}) would then be similar to those
of \cite{VCFH,CBV} although our velocity equation, which we derived
from the microscopic dynamics, is rather more complicated and contains
additional terms.

\section{Discussion}
\label{D}
In this work we have extended the flocking model of \cite{EOL} to
include all three of Reynolds' three effects---alignment, centring and
separation---implemented in combination or independently.  Within this
model we demonstrated the robustness of the alternating flock
highlighted in \cite{EOL} to the addition of other Reynolds' effects.
Also we showed the existence of two new regimes: the homogeneous flock
and dipole structures. These are consistent with results from a mean
field treatment which, in particular, correctly predicts the form of
the dipole solutions. Furthermore, we investigated the coarsening
process in the low density dipole regime.

We also derived from microscopic considerations the proposed form for
the function $G$ (\ref{abias}).  The algorithm presented in section
\ref{JOFFVCA} is not only intuitive but significantly is a fixed time
algorithm, by which it is meant that the algorithm would involve
sampling a fixed number of neighbouring boids whatever the
density. Such algorithms are aspired to in flocking applications
\cite{Rey,BDG} and are presumed to underly natural boid decision
making \cite{HBSM}. Several ways to extend the algorithm could be to
include a density, velocity or historical dependence in determining
the sample, anisotropic spatial sampling or considering a direction
selection rule other than majority. For example boids could require
unanimity\cite{Ray} amongst their observed neighbours to believe that
any orientation is a worthy choice.

Reynolds' three effects were defined as required behaviours for
simulating real flocks. In this paper we have shown several states
which may characterise real systems. Firstly the {\it homogeneous
flock}, which has many analogies in two or three dimensions, can be
realised in some systems approximating one dimension: for example
skaters confined to the outside of a rink, or fish in a ring shaped
tank (as in the Boston Aquarium). The {\it fixed density dipole} is
characterised by
sharp edges and fixed
internal density, and the ability to wander---
either slowly, or with the inclusion of alignment, quickly and with
a sustained orientation. These qualities, as well as
as the ability to split from within,
could describe a number of biological systems. Finally
consider the reversal mechanism present in the alternating
flock; such sharp reversals of direction are characteristic of the
manoeuvres seen in many flocks, though they are poorly described by
many models.

The numerical results obtained for the full stochastic system have
some noteworthy similarities with previous studies. 
A flocking state
is observed in the one dimensional continuous space model of
\cite{CBV} which appears to resemble an alternating flock.
Dipole-like regimes have been observed in
\cite{LR}, where dense states are supported by a centring
interaction. As with our dipoles the density at the centre of these
systems grows rapidly with boid number. This effect is suppressed by a
hard core repulsion, producing states similar to our fixed density
dipoles. In \cite{LR} two dimensional systems also display wandering
and oscillating (circling) behaviour. In \cite{BDG} a two dimensional
cellular automaton demonstrates a density-dependent state with 
symmetry-broken velocity. This density dependence arises from a {\it hard}
capacity being placed on the number of boids which can occupy a single
site; in effect the flocking state is destroyed by separation at high
density.

Our mean field treatment has similarities with the continuum model in
\cite{CBV}. Not only are the same states demonstrated, but the terms
within the equations are comparable also. Special note is made in
\cite{VCFH} of a term of the form
$\frac{1}{\rho}\frac{\partial}{\partial x} G \frac{\partial}{\partial
x} \rho$ which is present in our equation (\ref{continuumv}). This
term relates to domain competition, and aids in an understanding of
the reversal mechanism and breakdown of the homogeneous flock. 
As (\ref{continuumv}) describes very well the shape of a spreading flock,
and it is believed that this shape allows certain fluctuations
to cause the flock reversal mechanism \cite{EOL},
it might be hoped that the addition of noise into
(\ref{continuumrho},\ref{continuumv}) would allow the alternating
flock to be further analysed.  
The coarsening process in low and high density dipole regimes
may also be observed by allowing fluctuations to enter the dynamics at
the boundaries of the dipoles.

Finally we mention one further variation of the model of \cite{EOL} that we
have  studied \cite{Ray}.
To imitate the effects of inertia we modified step (iii)
of our dynamics such that boids which decide to change direction must
proceed through a temporary zero velocity state. Numerical simulations
and a mean field treatment  (which we do not present here)
indicate that under such a scheme flocking
solutions become more prominent and have higher global velocities. 
Although the flipping mechanism proceeds in much the same manner as
the two velocity case,
the rate of reversal for both the alternating and homogeneous flock is
reduced This is not surprising since more
unfavoured flips are required to create momentum fluctuation leading
to  a sustainable reversal of the velocity.\\

\noindent
{\bf Acknowledgements} 
We thank Mike Cates and Jamie Wood for carefully reading the manuscript and for helpful
discussions. The work was supported in part by EPSRC Programme grant GR/S10377/01.

\appendix
\section {Derivation of Mean Field Equations}
\label{appendixa}

In our mean field theory we
shall approximate the density $\rho(x,t)$ (the number of particles at site $x$ and time $t$)
and  the momentum $\phi(x, t)$  (the number of right going minus the number of left moving particles 
at site $x$ and time $t$) by continuous functions and derive the evolution
of these funcions by ignoring certain correlations.
We  begin with the dynamics of site $x$ in a
single update, as defined by $\rho(x ,t)$ and $\phi(x , t)$. Consider
$\rho(x,t)$, in a single update this can increase if a boid selected
from a neighbouring site moves into site $x$, or can decrease if a
boid is selected from site $x$, with an associated probability:
\begin{eqnarray}
\lefteqn{ \rho(x , t+\delta t)= \hspace{0.5cm} }\\
&& \left\{\begin{array}{ll}\rho(x , t)+1,&
  \mbox{Probability}  = \frac{\rho(x-a , t)}{N}W_+(x-a , t) +
  \frac{\rho(x + a , t)}{N}W_-(x + a , t)\nonumber \\ \rho(x , t) -1,&
  \mbox{Probability } =  \frac{\rho(x , t)}{N}\nonumber \\ \rho(x , t),&
  \mbox{ otherwise.}\end{array}\right.
\end{eqnarray}

We then average over the events
occurring between time $t$ and $ t+ \delta t$
 \begin{eqnarray}
\rho(x , t+\delta t) &=& \sum_{\rho(x , t+\delta t)} \rho(x , t+\delta t) P_{\rho(x , t+ \delta t)} 
\end{eqnarray}
and make the mean field approximation of factorising all averages.

Further we assume $\rho(x,t)$ and $\phi(x,t)$ vary smoothly on the
time and length scales effective in a single update. With these
assumptions we expand about site $x$ to second order (in lattice
spacing $a$) and about $t$ to first order ($\frac{1}{N}$).
\begin{eqnarray}
 \frac{\partial \rho}{\partial t} &=& \lbrace \rho - a \frac{\partial \rho}{\partial x} + \frac{a^2}{2} \frac{\partial^2 \rho}{\partial x^2}\rbrace\lbrace W_+ - a \frac{\partial W_+}{\partial x} + \frac{a^2}{2}\frac{\partial^2 W_+}{\partial x^2}\rbrace  \nonumber \\
  &+& \lbrace \rho + a \frac{\partial \rho}{\partial x} + \frac{a^2}{2} \frac{\partial^2 \rho}{\partial x^2}\rbrace\lbrace W_- + a \frac{\partial W_-}{\partial x} + \frac{a^2}{2}\frac{\partial^2 W_-}{\partial x^2}\rbrace  \nonumber \\
  &-& \rho  \;.
\end{eqnarray}

By introducing $G$ (\ref{G}) the following equation can be
determined. In this equation there is a diffusion term arising from
the choice of random stochastic updates, and a current  term
controlled by the mass current  $\rho G$.
\begin{eqnarray}
  \frac{\partial \rho}{\partial t} &=& -a \frac{\partial}{\partial x} (\rho G) + \frac{a^2}{2}\frac{\partial^2}{\partial x^2} \rho\; . \label{rhoeq}
\end{eqnarray}
In a similar way one can develop an equation for $\phi(x,t)$ beginning with the full dynamics at $x,t$, 
\begin{equation}
  \phi(x , t+\delta t) = \left\{\begin{array}{ll}\phi(x , t)+1,& \mbox{ Probability = }\frac{\rho(x-a , t)}{N}W_+(x-a , t) + \frac{n_{-1}(x , t)}{N}\nonumber \\
  \phi(x , t)-1,& \mbox{ Probability = } \frac{\rho(x + a , t)}{N}W_-(x + a , t) + \frac{n_1(x , t)}{N}\nonumber \\
  \phi(x , t),& \mbox{ otherwise.}\end{array}\right.
\end{equation}
Taking the expectation value and expanding to second order in $a$ and first order in $t$:
\begin{eqnarray}
  \frac{\partial \phi}{\partial t} &=& \lbrace \rho - a \frac{\partial \rho}{\partial x}+ \frac{a^2}{2}\frac{\partial^2 \rho}{\partial x^2}\rbrace\lbrace W_+ - a \frac{\partial  W_+}{\partial x} + \frac{a^2}{2} \frac{\partial^2 W_+}{\partial x^2}\rbrace   \nonumber \\
  &-& \lbrace \rho + a\frac{\partial  \rho}{\partial x} + \frac{a^2}{2}\frac{\partial^2 \rho}{\partial x^2}\rbrace\lbrace W_- + a \frac{\partial W_-}{\partial x} + \frac{a^2}{2} \frac{\partial^2 W_-}{\partial x^2}\rbrace   \nonumber \\
  &-& \phi  
\end{eqnarray}
Using (\ref{G}) this can be simplified to its final form
\begin{eqnarray}
  \frac{\partial \phi}{\partial t} &=& -\phi + \rho G - a \frac{\partial}{\partial x} \rho +\frac{a^2}{2} \frac{\partial^2}{\partial x^2} ( \rho G) \label{phieq}\;.
\end{eqnarray}
Defining  the (neighbourhood) velocity as
\begin{equation}
v(x,t) = \frac{\phi(x,t)}{\rho(x,t)}\;.
\end{equation}
yields using (\ref{rhoeq},\ref{phieq}) a velocity equation:
\begin{eqnarray}
  \frac{\partial v}{\partial t} &=&
 -v + G + \frac{a}{\rho}(v\frac{\partial}{\partial x} (\rho G)-\frac{\partial}{\partial x} \rho) + \frac{a^2}{2\rho}(- v\frac{\partial^2}{\partial x^2} \rho+\frac{\partial^2}{\partial x^2} (\rho G))\;.  \nonumber
\end{eqnarray} 

We now  write   $G(x,t)$, the expected update velocity, as a function of $v(x,t)$ and $\rho(x,t)$ only.  To simplify we keep only  leading order terms  in $a$ and obtain
\begin{equation}
  G = (1 - 2 \eta)\frac{\tanh\left( \beta \left[ (1-\kappa) v + \kappa \omega \right]\right) }{\tanh\beta} 
\end{equation}
where 
\begin{eqnarray}
  \omega &=& \frac{2 a }{3 \rho}\ \frac{ \partial \rho(x,t)}{\partial x}\;.
\label{omegaa}
\end{eqnarray}
Finally, setting $a$ to one 
in (\ref{rhoeq},\ref{phieq}, \ref{omegaa})
yields (\ref{continuumrho}--\ref{OmegaMF}).

\end{document}